\documentclass[letterpaper, 10pt, oneside, conference, final]{IEEEtran}
\usepackage[latin1]{inputenc}
\usepackage[pdftex]{graphicx}
\usepackage{array, multirow}
\usepackage{color}
\usepackage{booktabs}
\usepackage{caption}
\usepackage{subcaption}
\usepackage{tikz, pgfplots}
\usetikzlibrary{arrows.meta}
\usetikzlibrary{spy}
\usepackage{amssymb}
\usepackage{amsmath}
\usepackage{tabularx}
\usepackage{makecell}
\usepackage[top=0.742in, bottom=1.086in, left=0.68in, right=0.68in]{geometry}
\pgfplotsset{compat=1.16}
\def\BibTeX{{\rm B\kern-.05em{\sc i\kern-.025em b}\kern-.08em
		T\kern-.1667em\lower.7ex\hbox{E}\kern-.125emX}}

\usepackage[nolist]{acronym}
\begin{acronym}   
	\acro{DL}{deep learning}    
	\acro{DNN}{deep neural network} 
	\acro{LS}{least squares}   
	\acro{MLE}{maximum likelihood estimator}
	\acro{MSE}{mean squared error}
	\acro{RMSE}{root mean squared error}
	\acro{WSN}{wireless sensor network}
	\acro{RSS}{received signal strength}
	\acro{IoT}{Internet of Things}
	\acro{5G}{fifth generation}
	\acro{SU}{sensing unit}
	\acro{CU}{central unit}
	\acro{CRLB}{Cram\'{e}r-Rao lower bound}
	\acro{MPE}{mean position error}
	\acro{CDF}{cumulative distribution function}
	\acro{ToA}{time of arrival}
	\acro{AoA}{angle of arrival}
	\acro{REML}{radio enviroment map localization}
	\acro{OWL}{online wireless lab}
	\acro{AGV}{automated guided vehicle}
	\acro{SDR}{software-defined radio}
	\acro{OWL}{Online Wireless Lab}
	\acro{PLM}{path loss model}
	\acro{ELU}{exponential linear unit}
\end{acronym}

\begin{document}
\IEEEoverridecommandlockouts
\title{Experimental Performance of Blind Position Estimation Using Deep Learning}
\author{
	\IEEEauthorblockN{Ivo Bizon\IEEEauthorrefmark{1}, Zhongju Li\IEEEauthorrefmark{1}, Ahmad Nimr\IEEEauthorrefmark{1}, Marwa Chafii\IEEEauthorrefmark{2}\IEEEauthorrefmark{3} and Gerhard P. Fettweis\IEEEauthorrefmark{1}}
	\IEEEauthorblockA{\IEEEauthorrefmark{1}Vodafone Chair Mobile Communications Systems, Technische Universit{\"a}t Dresden (TUD), Germany}
	\IEEEauthorblockA{\{ivo.bizon, zhongju.li, ahmad.nimr, gerhard.fettweis\}@ifn.et.tu-dresden.de}
	\IEEEauthorblockA{\IEEEauthorrefmark{2}Engineering Division, New York University (NYU), Abu Dhabi, UAE}
	\IEEEauthorblockA{\IEEEauthorrefmark{3}NYU WIRELESS, NYU Tandon School of Engineering, New York, USA}
	\IEEEauthorblockA{marwa.chafii@nyu.edu}
}
\maketitle
\begin{abstract}
	Accurate indoor positioning for wireless communication systems represents an important step towards enhanced reliability and security, which are crucial aspects for realizing Industry 4.0.
	In this context, this paper presents an investigation on the real-world indoor positioning performance that can be obtained using a \ac{DL}-based technique.
	For obtaining experimental data, we collect power measurements associated with reference positions using a wireless sensor network in an indoor scenario. 
	%	Part of this collected data is used for training and part is used for evaluating the positioning performance. 
	The \ac{DL}-based positioning scheme is modeled as a supervised learning problem, where the function that describes the relation between measured signal power values and their corresponding transmitter coordinates is approximated.
	We compare the \ac{DL} approach to two different schemes with varying degrees of online computational complexity. Namely, maximum likelihood estimation and proximity. 
	Furthermore, we provide a performance comparison of \ac{DL} positioning trained with data generated exclusively based on a statistical path loss model and tested with experimental data.
\end{abstract}
\begin{IEEEkeywords}
	Blind localization, wireless sensor network, received signal strength, deep learning, positioning.
\end{IEEEkeywords}
\acresetall

\section{Introduction} \label{introduction}
\IEEEPARstart{T}{he} growing interest for accurate indoor positioning\footnote{The terms \emph{location} and \emph{position} are used interchangeably in this paper, and they refer to the Cartesian coordinates of one or more active transmitters within an area of interest, i.e., local positioning.} techniques can be attributed to the increasing number of envisioned use cases for future wireless networks. 
As an example, enhanced emergency calls will have to provide accurate positioning services indoors for aiding rescue teams \cite{FCC_indoor_loc_req}.
Similarly, proactive and privacy preserving localization of intentional or accidental interfering transmitters is a requirement for guaranteeing reliability and security in private wireless networks \cite{comms_in_6g_era}.
Both scenarios will benefit from positioning schemes that are able to operate with accuracy in the order of 1-5 meters in harsh propagation conditions, and with minimal prior knowledge of the transmission protocol and propagation characteristics.
% Other examples: Dynamic radio resource management. Warehouse automation employing \acp{AGV}. Real-time spectrum and interference monitoring.
Therefore, indoor positioning/localization enabled by \ac{DL} is expected to play a key role in future wireless networks \cite{6g, pos_loc_futuristic_comm}.
However, \ac{DL} requires access to large amounts of labeled data to achieve the required performance, and this can hinder its widespread adoption.
%This has motivated our investigation on reducing the dependency of measurement data by providing synthetic data for training the \ac{DL} model.

Among the possible location-dependent signal parameters, \ac{RSS} and \ac{AoA} allow blind localization, i.e., without previous knowledge of the transmitting protocol.
For the latter case, an antenna array is required for measuring the \ac{AoA} at each \ac{SU}. However, this  translates to higher implementation costs when compared to the hardware needed for \ac{RSS} measurement.
Moreover, the algorithms that offer high precision \ac{AoA} estimation often present complexity levels that hinder real-time implementation \cite{handbook_ch9}.
Thus, it can be argued that \ac{RSS} is the most convenient source of information for blind positioning.

In \cite{blind_tx_loc} we propose a blind transmitter localization framework using a \ac{DNN}, and investigate its performance when compared to classical and state-of-the-art approaches.
In this paper, we apply the approach presented in \cite{blind_tx_loc} to analyze the performance of \ac{DL}-based positioning when real-world \ac{RSS} measurements \cite{dataset} are available, and to understand the achievable real-time performance by employing such technique.
Furthermore, due to the significant cost associated with measurement campaigns, the localization performance is also investigated assuming synthetic data generation. The data are obtained through well accepted mathematical models that describe the relationship between the transmitters' physical location and the corresponding \ac{RSS}, i.e., statistical \acp{PLM}.
This approach  leverages theoretical knowledge to reduce, or completely eliminate, the need for real-world measurements.

Our contributions can be summarized as follows: 
\begin{enumerate}
	\item We validate by means of real-world measurements the \ac{DL} localization approach presented in \cite{blind_tx_loc}.
	
	\item We analyze the performance of our \ac{DL} localization scheme when the training data are synthetically generated with a \ac{PLM}. This provides insights into the localization performance that can be achieved when no measured data are available, and only theoretical knowledge on propagation is used.
	
	\item Our \ac{DL} scheme is compared with other solutions of different computational complexity, namely, \ac{MLE} and Proximity. The observed performance suggests that \ac{DL} can provide accurate positioning using \ac{RSS} as a source of position information in indoor scenarios.
\end{enumerate}

The remainder of the paper is organized as follows: 
Section \ref{sec:measurement setup} presents the measurement setup used for data collection.
Section \ref{sec:plm} presents the log-normal PLM used for fitting the measured data and generating synthetic training data.
Section \ref{sec:localization techniques} describes the localization algorithms.
Section \ref{sec:results} analyses the performance of the localization schemes. 
Finally, the paper is concluded in Section \ref{sec:conclusions}.
\section{Measurement Setup and \ac{RSS} Calculation} \label{sec:measurement setup}

\begin{figure*}[t]
	\centering
	\begin{subfigure}[b]{0.3\textwidth}
		\centering
		\includegraphics[width=0.87\textwidth]{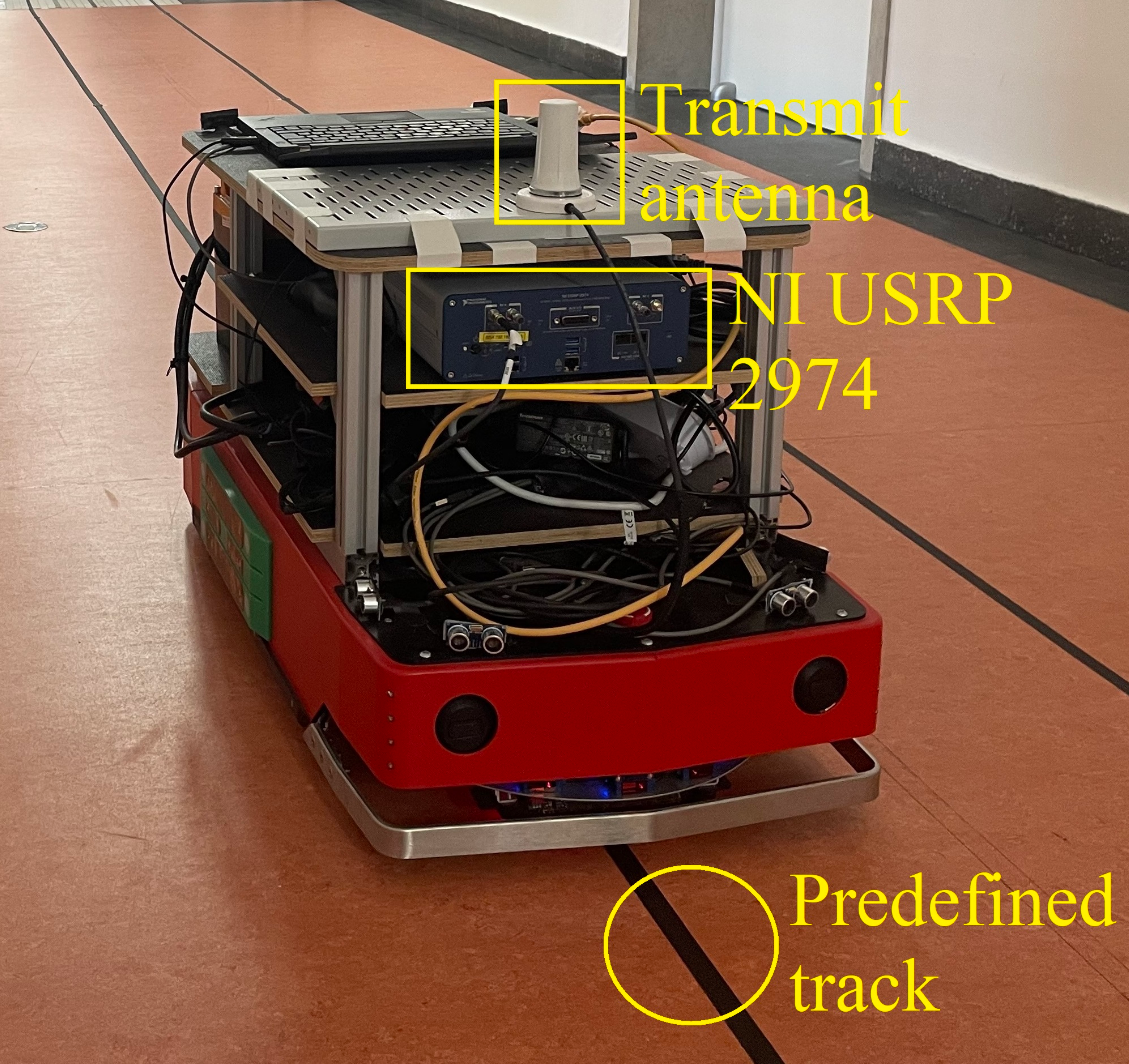}
		\caption{Probing transmitter on the AGV.}
		\label{agv}
	\end{subfigure}
	\hfill
	\begin{subfigure}[b]{0.32\textwidth}
		\centering
		\includegraphics[width=\textwidth]{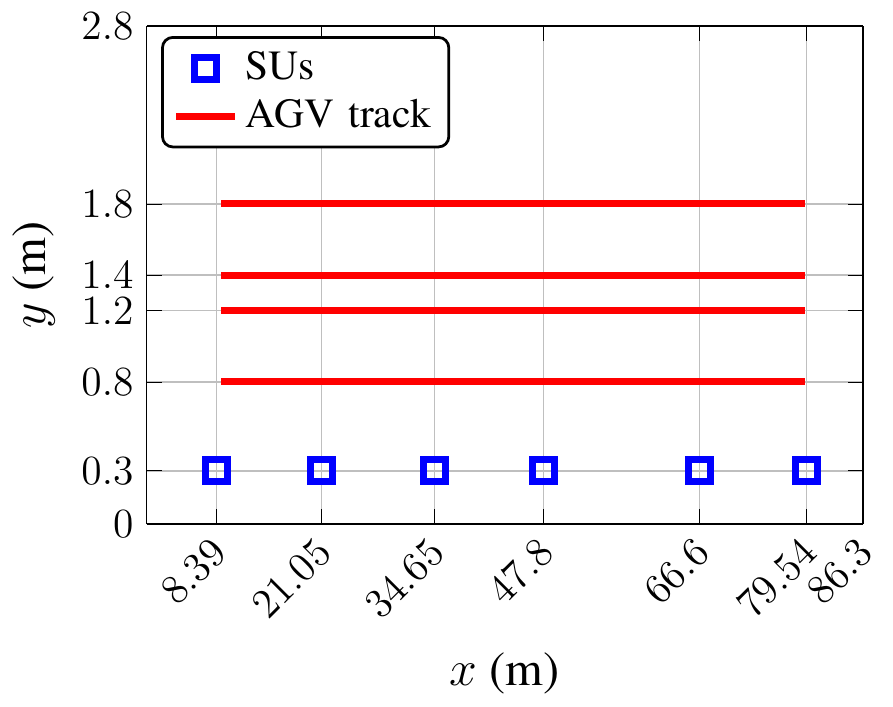}
		\caption{Illustration of the measurement area.}
		\label{corridor}	
	\end{subfigure}
	\hfill
	\begin{subfigure}[b]{0.3\textwidth}
		\centering
		\includegraphics[width=0.87\textwidth]{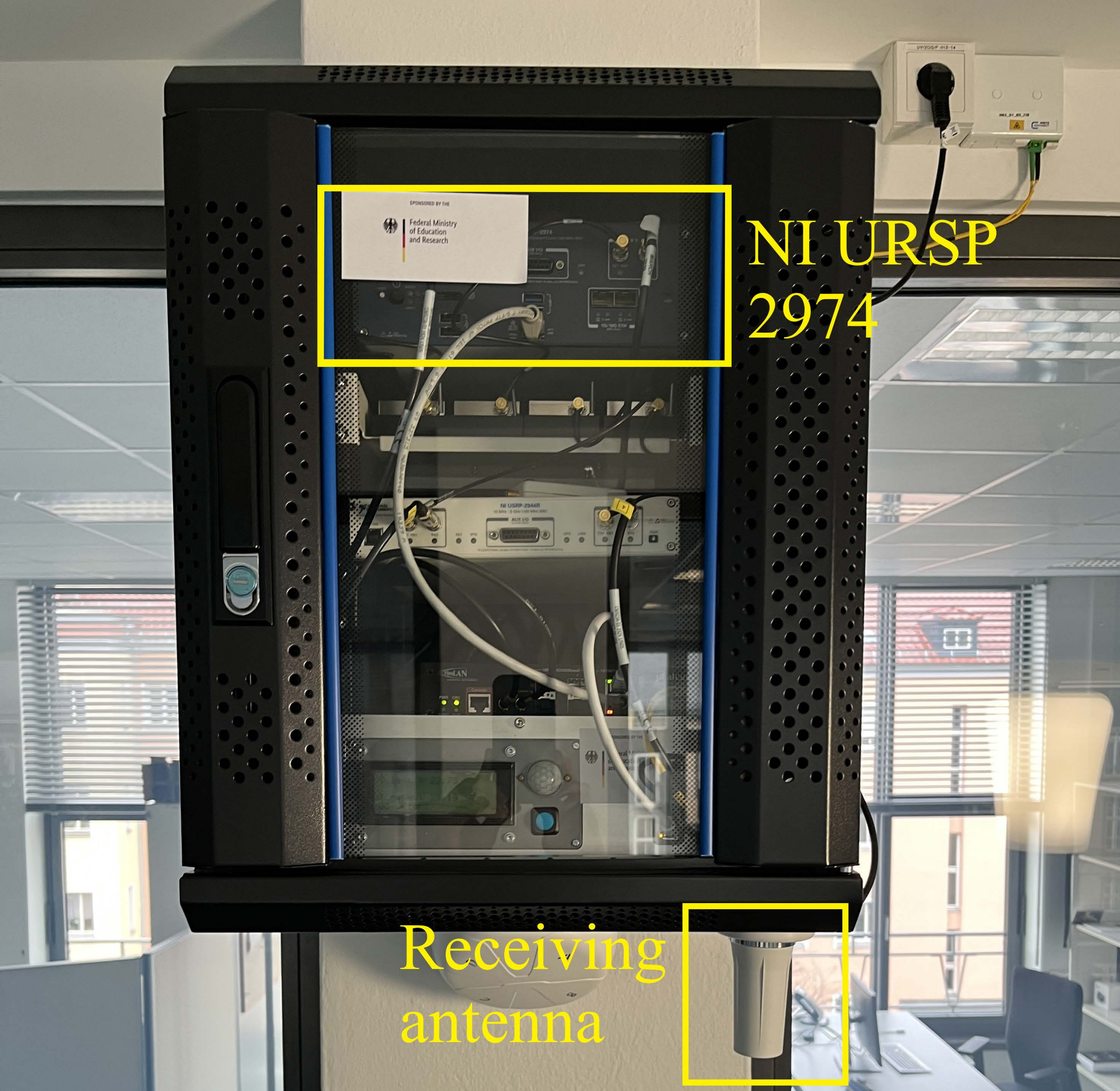}
		\caption{One of the 6 sensing units.}
		\label{su}
	\end{subfigure}
	\caption{Components employed in the measurement campaign.}
	\label{measurement_scenario}
\end{figure*}

The measurements are conducted using the hardware resources from the indoor \ac{OWL} testbed located at the Technische Universit{\"a}t Dresden (TUD), Dresden, Germany \cite{owl}.
The indoor \ac{OWL} has 6 access points mounted on the walls of an office corridor with antennas at a height of 2.5 m.
These access points are configured as the \acp{SU}.
A probing transmitter is carried by an \ac{AGV} that follows a predefined track with the constant speed of 0.6 m/s.
The transmit antenna is 0.5 m above the ground. 
Fig. \ref{measurement_scenario} shows the components used in the measurement campaign with an illustration of the area where the measurements are conducted along the predefined track followed by the \ac{AGV}, and the coordinates of the \acp{SU} are also shown.
Both probing transmitter and \acp{SU} are implemented using the Universal Software Radio Peripheral (USRP), which is a \ac{SDR} platform by National Instruments.
The center frequency of operation of the radios is 3.75 GHz.
The measurement environment is an indoor office hallway of size 86.3 m $\times$ 2.8 m.
We define one measurement round as the \ac{AGV} moves from the initial measurement point with $x_{\mathrm{init}} = 9$ m until $x_{\mathrm{end}} = 79.32$ m and back.
The $y$ coordinates of the transmit antenna vary from 0.8, 1.2, 1.4 and 1.8 m as we place it on the right or left edges of the carrying platform on the \ac{AGV} for different rounds.
This is illustrated in Fig.~\ref{corridor}.

The probing transmitter continuously sends a constant envelop chirp signal with 512 samples occupying a bandwidth of 100 MHz. 
The transmit power is set to 13 dBm. 
All \acp{SU} collect 5000 IQ samples at the rate of 100 Msample/s, such that one measurement period lasts 50 $\mu$s, and it is repeated every 200 ms. 
Consequently, the measurements are spaced apart by 12 cm.
Since the \ac{AGV} runs with a constant speed of 0.6 m/s, during one measurement period it moves 30 $\mu$m.
Therefore, we can treat it as static within one measurement period.
%One measurement is performed every 200 ms. 

In total, 4 measurement rounds have been performed during different times of the day, resulting in 4696 pairs of reference positions and the corresponding received IQ-samples.
From the total pairs, 75\% (3522 pairs) are reserved for training the \ac{DL} model, and the remaining 25\% (1174 pairs) are used as testing data.
Before the split, the data are shuffled to ensure statistical reliability.

At each \ac{SU}, the received IQ samples are used for calculating the \ac{RSS} as follows, 
\begin{equation}
	P_j = 10 \log_{10}\left( \frac{1}{N_y} \sum_{n=0}^{N_y-1} |y[n]|^2 \right) + 30,
\end{equation}
where $P_j$ (dBm) is the measured \ac{RSS} at the $j$-th \ac{SU}, and $y[n]$ represents the received signal with $N_y = 5000$ samples. %The superscript "m" denotes measurement.
\section{Path Loss and Shadowing Modeling} \label{sec:plm}
To generate the synthetic training data, the single slope log-normal \ac{PLM} \cite{rappaport} has been employed. 
Under this model, the \ac{RSS} measurement obtained at the $j$-th \ac{SU} is given by
\begin{equation} \label{lognormal}
	P_{j} =  P_0 - 10\beta\log_{10}\left(\frac{d\left(\mathbf{u}, \mathbf{v}_j\right)}{d_0}\right) + n_j,
\end{equation}
where $P_0$ (dBm) is the received power at a reference distance $d_0$, which includes the transmit power, and the transmit and receive antenna gains, $\beta$ is the environment dependent path loss exponent, $n_j$ represents the shadowing noise, $\mathbf{u} \triangleq \left[u_{x}, u_{y}, u_{z}\right]^{\mathrm{T}}$ and $\mathbf{v}_j \triangleq \left[v_{j_x}, v_{j_y}, v_{j_z}\right]^{\mathrm{T}}$ contain the transmitter and $j$-th \ac{SU} coordinates in three dimensional space, respectively, and $d\left(\mathbf{u}, \mathbf{v}_j\right)$ represents the Euclidean distance, defined as
\begin{equation} \label{euclidian_distance}
	\small{d\left(\mathbf{u}, \mathbf{v}_j\right) = \sqrt{(u_{x} - v_{x_j})^2 + (u_{y} - v_{y_j})^2 + (u_{z} - v_{z_j})^2}.}
\end{equation}
The shadowing noise experienced at the \acp{SU} is modeled by a zero-mean Gaussian random vector $\mathbf{n}$ with spatially dependent covariance matrix $\mathbf{C} \in \mathbb{R}^{N_s \times N_s}$.
This accounts for the effects of signal blockage by objects in the environment and multi-path propagation. 
Therefore, the correlation between shadowing noise depends on the distance between \acp{SU}.
Assuming an exponential correlation model \cite{correlated_shadowing}, the covariance matrix of $\mathbf{n}$ is described as 
\begin{equation} \label{corr_mat}
	\left[\mathbf{C}\right]_{a,b} = \sigma_{\mathrm{dB}}^2 \exp\left(-\frac{d\left(\mathbf{v}_a, \mathbf{v}_b\right)}{d_\mathrm{cor}} \right),
\end{equation}
where $d_\mathrm{cor}$ represents the decorrelation distance, which is assumed to be 1 meter \cite{decorrelation_distance}, and $\sigma_{\mathrm{dB}}^2$ the shadowing noise variance in dB.

The model described by \eqref{lognormal} can be modified to include an extra attenuation term dependent on the number of walls or floors between transmitter and receiver. 
We opt to not include this term, since our experiments were carried out in a corridor without hard partitions.
\section{\ac{RSS} based localization techniques} \label{sec:localization techniques}

Let us consider that within the area of interest, $N_s$ \acp{SU} are placed in fixed locations. 
The \acp{SU} measure \ac{RSS} values and send them to a \ac{CU}, where the transmitter coordinates are estimated.

\subsection{Deep learning framework for blind localization}

Given the data set with known transmitter positions and associated \ac{RSS} measurements at fixed \acp{SU} obtained from our measurement campaign, the localization problem is modeled as a supervised learning problem. 
The architecture of the \ac{DNN}-based transmitter localization scheme employed in this paper has been proposed in our previous work \cite{blind_tx_loc}.
This model uses the \ac{RSS} measurement vector in dB scale for numerical stability.
The training process consists of an iterative minimization of a predefined loss function between the estimated and known training examples. 
The \ac{MSE} is employed as loss function, since it yields the \ac{MLE} when the data are Gaussian \cite{deep_learning}.
Therefore, the learned weights of the \ac{DNN} represent an approximation of the \ac{MLE}, but with significantly lower online implementation complexity. %than grid search.
However, the offline training of the network weights is an additional step, which requires computational resources. 
Nevertheless, the resulting \ac{DNN} model accounts for hardware and environmental particularities of each area once training is completed.
Hence, the resulting model yields an estimator that is particularly suited to perform well within the area of interest. In particular, it is important to note that the estimators obtained via data driven approaches are naturally immune to system modeling simplifications and misconceptions, since they are obtained from data sets based on real-world measurements.
Moreover, to investigate the reliability of our numerical simulations, the performance of the \ac{DNN} model is also presented when the training and testing data are obtained by \eqref{lognormal}.

The hyper-parameters of the selected architecture are presented in Table \ref{table_hyperparameters}.
\begin{table}[t]
	\centering
	\caption{\ac{DNN} hyperparameters}
	\label{table_hyperparameters}
	\renewcommand{\arraystretch}{1.3}
	\begin{tabular}{ll}
		\toprule[0.9pt] 
		Hyperarameter & Value \\ 
		\midrule
		Number of hidden layers & 3 \\
		Number of hidden units per layer & 128 \\
		Number of training examples & 3522 \\
		Validation split & 80\% training, 20\% validation \\
		Mini batch size & 40 \\
		Regularization parameter & 0.01 \\ 
		Activation function of hidden units & \ac{ELU} \\ 	
		Epochs & 2000 \\	
		Early stop patience & 100 \\ 
		Optimizer & Adaptive moments (Adam) \cite{adam} \\ 
		Learning rate & 10$^{-4}$ \\
		Loss function & MSE \eqref{mse} \\
		Weight initialization & Xavier \cite{xavier} \\
		\bottomrule[0.9pt]	
	\end{tabular}
\end{table}
For locating the transmitters, the last layer is linear, and outputs the estimated transmitter coordinates, which are given by
\begin{equation}
	\mathbf{\hat{u}}_\mathrm{DNN} = \mathbf{W}_L^{\mathrm{T}}\mathbf{a}_{L-1} + \mathbf{b}_L,
\end{equation}
where $\mathbf{\hat{u}}_\mathrm{DNN} \triangleq \left[\hat{u}_{x}, \hat{u}_{y}, \hat{u}_{z}\right]^{\mathrm{T}}$.
The \ac{MSE} is used as loss function between the estimated and true transmitter coordinates with a regularization parameter, and it can be written as
\begin{equation} \label{mse}
	\mathcal{L}_{\mathrm{MSE}} = \frac{1}{m} \sum_{k = 1}^{m} \mathrm{MSE}^{(k)}\left(\mathbf{\hat{u}}, \mathbf{u}\right) + \frac{\lambda}{2m} \sum_{w \in \mathbb{W}} w^2,
\end{equation}
where $m$ is the total number of training examples in one training batch, $\lambda$ is the regularization parameter and $\mathbb{W}$ represents the set that contains all weights and biases from the network, and 
\begin{equation}
	\mathrm{MSE}\left(\mathbf{\hat{u}}, \mathbf{u}\right) = \frac{1}{2}\sum_{i = 1}^{2} \left(\left[\mathbf{\hat{u}}\right]_{i} - \left[\mathbf{u}\right]_{i}\right)^2.
\end{equation}
The activation of the $l$-th hidden layer is given by
\begin{equation}
	\mathbf{a}_l = \mathrm{ELU}\left(\mathbf{W}_l^{\mathrm{T}}\mathbf{a}_{l-1} + \mathbf{b}_l\right),
\end{equation}
where $\mathbf{W}_l \in \mathbb{R}^{N_u^{(l-1)} \times N_u^{(l)}}$ and $\mathbf{b}_l \in \mathbb{R}^{N_u^{(l)} \times 1}$ contain the weights and biases associated with the edges connecting the ($l-1$)-th to the $l$-th layer, respectively, $N_u^{(l)}$ represents the number of units in the $l$-th layer, and $\mathrm{ELU}\left(x\right)$ is the exponential linear unit \cite{elu}.

\subsection{Maximum likelihood estimation}

The \ac{MLE} is derived assuming \eqref{lognormal} as the system model.
Moreover, it is assumed that a single transmitter is present in the area of interest, the measurement vector is available at the \ac{CU}, and without loss of generality, the reference distance is one meter.
Hence, the \ac{MLE} for the transmitter coordinates is obtained as
\begin{equation} \label{MLE}
	\boldsymbol{\hat{\theta}}_{\mathrm{ML}} = \underset{\boldsymbol{\theta}_{\mathrm{ML}} \in \mathbb{S}}{\arg\min} \big(\mathbf{m} - \mathbf{f}\left(\boldsymbol{\theta}_{\mathrm{ML}}\right)\big)^{\mathrm{T}} \mathbf{C}^{-1} \big(\mathbf{m} - \mathbf{f}\left(\boldsymbol{\theta}_{\mathrm{ML}}\right)\big),
\end{equation}
where $\boldsymbol{\theta}_{\mathrm{ML}}  \triangleq \left[ \mathbf{u},\; P_0,\; \beta\right]^{\mathrm{T}}$ represents the unknown parameter vector, $\mathbb{S}$ the set of possible values each parameter can take, and the $j$-th entry of $\mathbf{f}\left(\boldsymbol{\theta}_{\mathrm{ML}}\right)$ is given by
\begin{equation}
	\left[\mathbf{f}\left(\boldsymbol{\theta}_{\mathrm{ML}}\right)\right]_j = P_0 - 10\beta\log_{10}\left(d\left(\mathbf{u}, \mathbf{v}_j\right)\right).
\end{equation}
It is important to notice that the \ac{MLE} does not have a closed-form solution.
Therefore, it is implemented as a grid search over $\mathbb{S}$. 
Complexity can be reduced by assuming knowledge of $\beta$ and $P_0$.
%Linear estimators such as the \ac{LS} are low complexity alternatives, but the localization performance is significantly inferior when compared to \ac{MLE}.

\subsection{Proximity}

Proximity is the simplest localization scheme considered in this paper. 
It estimates the transmitter position as equal to the coordinates of the \ac{SU} that measures the highest \ac{RSS}.
Due to its simple definition, its performance is used as a benchmark for comparison with the more sophisticated schemes.
\section{Results} \label{sec:results}

\subsection{Estimation of \ac{PLM} parameters}
\begin{figure}[t]
	\centering
	\includegraphics[]{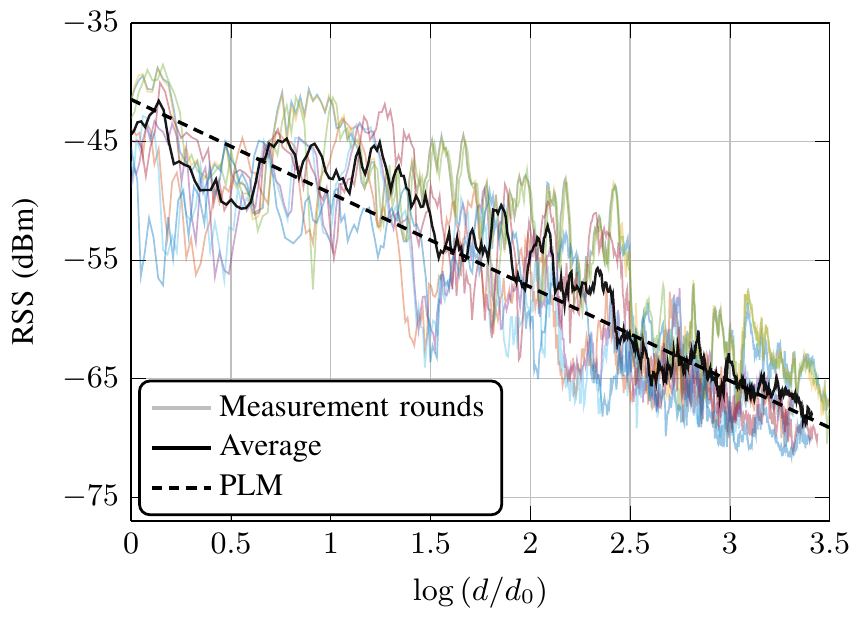}
	\caption{\ac{RSS} measurements and \ac{PLM} parameterized by $\hat{\beta}=1.82$.}
	\label{measurements}
\end{figure}
The data obtained from the measurement rounds contains a total of $L$ \ac{RSS} measurements at distinct coordinates, which are used for estimating the \ac{PLM} parameters assuming \eqref{lognormal} as model.
For this estimation task, \ac{LS} estimation is employed \cite{ls_plm_estimation}.

Ignoring the noise term, \eqref{lognormal} can be rewritten in a matrix form as
\begin{equation}
	\mathbf{y} = \mathbf{X}\boldsymbol{\theta},
\end{equation}
%$\mathbf{y} = \left[P(d_1/d_0), \, \cdots, \, P(d_L/d_0) \right]^{\mathrm{T}} \in \mathbb{R}^{L\times1}$
where $\boldsymbol{\theta} = \left[P_0, \, \beta \right]^{\mathrm{T}}$, $\mathbf{y} = \left[P(\mathbf{d}/d_0)\right]^{\mathrm{T}} \in \mathbb{R}^{L\times1}$ contains the \ac{RSS} measurements in dBm, $\mathbf{X} = \left[\mathbf{1}, \, -10\log_{10}(\mathbf{d}/d_0) \right] \in \mathbb{R}^{L\times2}$, where $\mathbf{1}$ is an $L$-long all ones vector, and $\mathbf{d}$ contains the $L$ distinct measurement distances.
Hence, the parameter vector is obtained as
\begin{equation}
	\boldsymbol{\hat{\theta}} = (\mathbf{X}^{\mathrm{T}}\mathbf{X})^{-1}\mathbf{X}^{\mathrm{T}}\mathbf{y},
\end{equation}
and the shadowing noise variance is given by
\begin{equation}
	\hat{\sigma}^2_\mathrm{dB} = \frac{1}{L-1}(\mathbf{y} - \mathbf{X}\boldsymbol{\hat{\theta}})^{\mathrm{T}}(\mathbf{y} - \mathbf{X}\boldsymbol{\hat{\theta}}).
\end{equation}

Fig.~\ref{measurements} presents the measured \ac{RSS} at the \ac{SU} located at (8.39, 0.3) for all 4 rounds. 
The corresponding averaged \ac{RSS}  and the fitted \ac{PLM} curves are also shown.
The estimated \ac{PLM} parameters are $\hat{\beta}=1.82$ and $\hat{\sigma}^2_\mathrm{dB}=11.83$, which agrees with similar measurement campaigns \cite{rappaport,experimental_rss_localization,PLE_experimental_analysis}. 

\subsection{Localization accuracy}

\begin{figure}[t]
	\centering
	\includegraphics[]{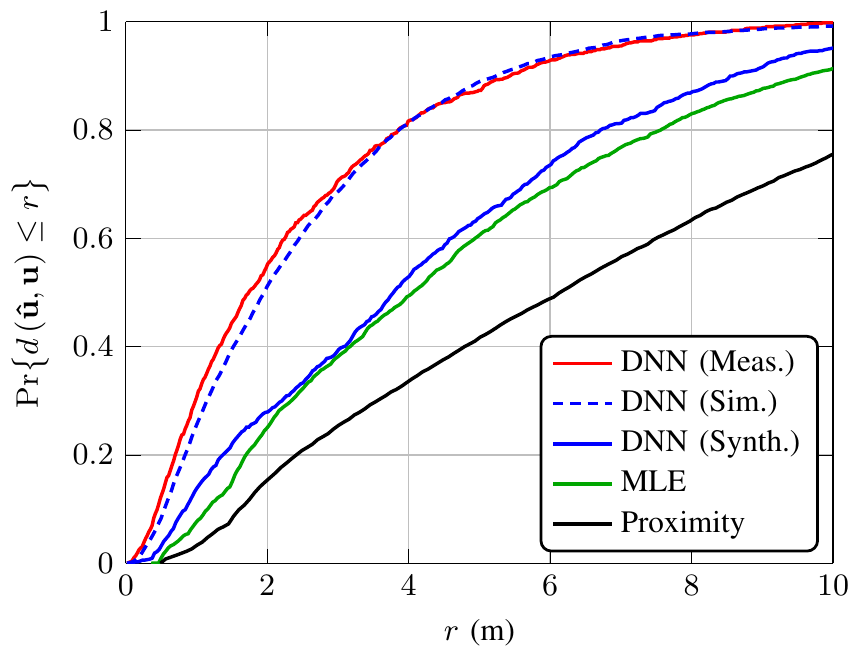}
	\caption{CDF of localization error. Solid lines are obtained from measured data saved exclusively for testing, and the dashed blue line employs synthetic data for testing.}
	\label{cdf_loc_err}
\end{figure}

\begin{table}[t]
	\centering
	\caption{Localization performance statistics}
	\label{localization_performance}
	\renewcommand{\arraystretch}{1.3}
	\begin{tabularx}{0.95\columnwidth}{lllll}
		\toprule[0.9pt] 
		%		Estimator & \makecell{Mean\\localization\\error (m)} & \makecell{Standard\\deviation (m)} & \makecell{Maximum\\localization\\error (m)} & \makecell{Minimum\\localization\\error (m)} \\ 
		Estimator & Mean LE & SD & Max. LE & Min. LE \\ 
		\midrule
		DNN (Meas.) & 2.3986 m & 2.0616 m & 11.6976 m & 0.0225 m \\
		DNN (Sim.) & 2.5153 m & 2.0986 m & 14.2105 m & 0.0151 m \\
		DNN (Synth.) & 4.3096 m & 3.0279 m & 16.7508 m & 0.0730 m \\
		MLE & 4.8009 m & 3.4373 m & 17.7379 m & 0.3734 m \\
		Proximity & 7.0752 m & 5.0404 m & 37.5100 m & 0.5000 m \\
		\bottomrule[0.9pt]	
		\multicolumn{4}{l}{\footnotesize{Localization error (LE), standard deviation (SD)}}
	\end{tabularx}
\end{table}

For assessing the localization performance, we present the \ac{CDF} of the localization error, and Table \ref{localization_performance} shows other statistics from the obtained results. 
The localization error is calculated using \eqref{euclidian_distance}.
Fig.~\ref{cdf_loc_err} shows the \ac{CDF} of the localization error for the presented schemes.
Solid lines are obtained from measured data saved exclusively for testing consisting of 1174 examples. 
The dashed line corresponds to the \ac{DNN} performance when synthetic data is used for training and testing, with the same amount of examples as obtained during the measurement campaign.
As we can observe, the performance of \ac{DNN} (Meas.) and \ac{DNN} (Sim.) are in agreement.
This result suggests that training and analyzing the performance of the \ac{DL} approach solely with simulated data gives meaningful insight about the expected performance in a real-world scenario.
Thus, corroborating the results presented in \cite{blind_tx_loc}. 
%However, this does not mean that the trained model with synthetic data can be directly used in realistic scenario. In particular, this can be observed from the performance gap between the DNN (Meas.) obtained by training with the  measurements,  and the model DNN (Synth.). obtained from the synthetic data, when both are tested with measurement data.

Fig. \ref{cdf_loc_err} also shows a performance gap between the \ac{DNN} model obtained from measurements, DNN (Meas.), and the model obtained from synthetic data, DNN (Synth.), when both are tested with measurement data.
Several reasons can be attributed to this gap.
For instance, the propagation parameters can vary depending on the transmitter position, and this phenomenon is not captured by the \ac{PLM} employed for synthetic data generation.
This non-homogeneity of the indoor propagation channel with respect to different locations has been noted in literature \cite{the_indoor_channel}.
Moreover, it has been shown that each \ac{SU} can add a random offset to the RSS measurement due to hardware characteristics \cite{best_pratice}.
Therefore, it becomes evident that \eqref{lognormal} does not capture all effects that characterize the relation between the transmitter position and measured \ac{RSS}. 
Another reason is the assumption of an ideal omnidirectional antenna by the \ac{PLM}, which is often not accurate in practice, since antennas present some level of directivity.
These phenomena are captured by the \ac{DNN} model when trained with measured data, which leads to the improved localization performance when compared with the model trained with synthetic data.
The \ac{MLE} presents similar performance to \ac{DNN} (Synth.), since both have been designed under the same \ac{PLM}, and \acp{DNN} can be understood as an approximation of \ac{MLE} \cite{deep_learning}.
The slight superior performance of \ac{DNN} (Synth.) over the \ac{MLE} can be attributed to the quantization error associated with the grid search implementation of the \ac{MLE}.
Moreover, \ac{MLE} and \ac{DNN} (Synth.) outperform the Proximity scheme, suggesting that if more accurate propagation models, such as ray tracing, are used for training data generation, the performance of \ac{DNN} (Synth.) could approach \ac{DNN} (Meas.).

\section{Conclusion} \label{sec:conclusions}

In this paper, we have shown that \ac{DL} based positioning using data from wireless sensor networks has the potential to be a key technique for providing accurate and reliable position information for improving services in future wireless communications systems. 
Our experimental investigation suggests that analysing the positioning performance of \ac{DL} schemes using synthetic data obtained from \acp{PLM} gives meaningful predictions of the achievable performance observed in real environments.
Furthermore, the potential and downsides of employing synthetic data for training \ac{DL} models has been discussed. 
The results suggest that simple \acp{PLM}, such as the single slope log-normal, can only provide limited performance, and generating training data through more accurate \acp{PLM} is required for reducing the dependency of \ac{DL} on site-specific measurements.
\section*{Acknowledgment}
This work was supported by the European Union's Horizon 2020 research and innovation programme through the project iNGENIOUS under grant agreement 957216, by the German Research Foundation (DFG, Deutsche Forschungsgemeinschaft) as part of Germany's Excellence Strategy - EXC 2050/1 - Project ID 390696704 - Cluster of Excellence "Centre for Tactile Internet with Human-in-the-Loop" (CeTI) and by the German Federal Ministry of Education and Research (BMBF) (AI4Mobile under grant 16KIS1177, and 6G-life under grant 16KISK001K). We also thank the Center for Information Services and High Performance Computing (ZIH) at TU Dresden.
\bibliographystyle{ieeetr}
\bibliography{my_references}
\end{document}